\input{epsf}
\documentstyle[aps,preprint]{revtex}
\begin{document}
\tightenlines

\title{On singular probability densities generated by extremal dynamics}

\author{Guilherme J. M. Garcia$^*$ and Ronald Dickman$^\dagger$\\
 {\small Departamento de F\'\i sica, Instituto de Ci\^encias Exatas}\\
 {\small Universidade Federal de Minas Gerais, Caixa Postal 702}\\
  {\small CEP 30123-970, Belo Horizonte - Minas Gerais, Brazil} }

\date{\today}

\maketitle
\vskip 0.5truecm

\begin{abstract}
Extremal dynamics is the mechanism that drives the Bak-Sneppen model
into a (self-organized) critical state, marked by a singular stationary
probability density $p(x)$.
With the aim of understanding of this phenomenon, 
we study the BS model and several
variants via mean-field theory and simulation. 
In all cases, we find that $p(x)$ is singular at one or more points,
as a consequence of extremal dynamics. Furthermore we show that
the extremal barrier $x_i$ always belongs to the `prohibited' interval, 
in which $p(x)=0$. Our simulations indicate that the Bak-Sneppen
universality class is robust with regard to changes in the updating 
rule: we find the same value for the exponent $\pi$ for all
variants. Mean-field theory, which furnishes an exact description
for the model on a complete graph, reproduces the character of the
probability distribution found in simulations.  For the modified processes
mean-field theory takes the form of a functional equation for $p(x)$.
\vspace{1em}

\vspace{1em}

\noindent keywords: extremal dynamics; Bak-Sneppen model;
mean-field theory; functional equations; universality
\vspace{1em}

\noindent$^*$ Electronic address: gjmg@fisica.ufmg.br\\
$^\dagger$ Electronic address: dickman@fisica.ufmg.br\\

\noindent Corresponding author: Ronald Dickman\\
telephone: 55-31-3499-5665\\
FAX: 55-31-3499-5600

\end{abstract}

\newpage

\section{Introduction}

The Bak-Sneppen (BS) model was proposed as a possible explanation of 
mass extinctions observed in the fossil record \cite{Bak and Sneppen:1993},
and was recently adapted to model experimental data on bacterial populations 
\cite{Donangelo and Fort:2002,Bose:2001}. Independent of its biological
interpretation, the model has atracted much attention as a prototype of
self-organized criticality (SOC) under extremal dynamics \cite{Dickman et al:2000,head}.
The model has been studied through various approaches,
including simulation \cite{Grassberger:1995,Rios:1998,Boettcher:2000},
theoretical analysis \cite{Meester:2002,Li and Cai:2000,Dorogovtsev:2000},
probabilistic analysis (run time statistics) 
\cite{Caldarelli:2002,Felici:2001},
renormalization group \cite{Marsili:1994,Mikeska:1997},
field theory \cite{Paczuski:1994}
and mean-field theory \cite{Dickman et al:2000,head,Pismak:1997}.
Some variants have been proposed, for example the anisotropic BS model
\cite{Head:1998,Maslov:1998}. In this paper, we study the consequences
of extremal dynamics, using mean-field theory and simulation. With this aim,
we propose variants of the model and analyse how varying the
updating rule affects the stationary probability density and the critical
behaviour. 

In the evolutionary interpretation of the BS model, each site $i$
represents a ``niche" occupied by 
a single species, and bears a real-valued variable $x_i$ representing
the ``fitness" of this species.  (In the present context ``fitness"
denotes a propensity to resist extinction: if $x_i > x_j$ then 
species $j$ goes extinct before $i$, so that $x_i$ may be termed a
``barrier to extinction".)  
At each step the site with the smallest $x_i$, 
and its nearest neighbors, are replaced with randomly chosen values.
The replacement of the neighboring variables  
with new random values may be interpreted as a sudden unpredictable
change in fitness when a nearby niche (which might have borne a 
predator, or a food source for the species in question), is suddenly
colonized by a new species.  Selection, at each step, of the global
minimum of the $\{x_i\}$ (``extremal dynamics") represents a
highly nonlocal process, and would appear to require an external 
agent with complete information regarding the state of the system at each
moment.  Applications of the model in evolution
studies are reviewed in \cite{drossel02}.

In the original updating rule, neighbors are randomly affected by
the extinction of an interacting species. There is no {\sl a priori}
reason to expect that evolution should obey this specific rule on a
specific lattice. Thus, we ask: what happens if the extinction of the
least adapted species favors (or prevents) the extinction of 
other species? Moreover, what is the signature of extremal
dynamics that can, in principle, allow us to recognize it in the
real world?

Due to extremal dynamics, the BS model exhibits scale invariance in
the stationary state, in which
several quantites display power-law behaviour
\cite{Bak and Sneppen:1993}. Simulations show that
the stationary distribution of barriers follows a step function, 
being zero (in the infinite-size limit) for $x < x^* \simeq 0.66702(8)$
\cite{Grassberger:1995}.  Relaxing the extremal condition leads to a
smooth probability density and loss of scale invariance 
\cite{Dickman et al:2000,head}.

A striking feature of the BS model is that a simple updating rule leads to 
a singular stationary probability density $p(x)$.  An intriguing
question therefore arises, as to how changes in the updating rule
affect this density, an issue that has not, to our knowledge, been
investigated previously.  In this work we examine the consequences
of rules in which one or more sites are updated according to
$x \to x' = f(x)$, instead of being replaced by a random number.
(Here $f$ maps the interval [0,1] to itself.)  We find that this
can provoke dramatic changes in the stationary probability density.
In the extremal dynamics limit, the variant models belong to the same
universality class as the original.
We find that the hallmarks of extremal dynamics are that 
i) the stationary probability density is singular, and ii) with probability one,
the extremal $x_i$ (i.e., the next variable to be updated)
belongs to the `prohibited' region in which $p(x)=0$. 
Using a two-site mean-field approximation, we also find evidence that
nontrivial correlations are restricted to the prohibited region.

The balance of this paper is organized as
follows. Section II introduces the models, which are then analyzed using
mean-field like approaches in Sec. III.  In Sec. IV we present simulation
results, and summarize our conclusions in Sec. V.

\section{Models}

The Bak-Sneppen  
\cite{Bak and Sneppen:1993} model is a discrete-time Markov process
on a d-dimensional lattice of $L^d$ sites, with periodic boundaries. 
At each site we define a real-valued
variable $x_i(0)$.  Initially, these variables are independently
assigned random values uniform on [0,1). 
At time 1, the site $m$ bearing the minimum of all the numbers
$\{x_i(0) \}$ is identified, and it, along with its $2d$ nearest neighbors,
are given new random values, again drawn independently from
the interval [0,1). (In the one dimensional case considered here
this amounts to: $x_m(1) = \eta$,
$x_{m + 1}(1) = \eta'$, and $x_{m - 1}(1) = \eta''$,
where $\eta$, $\eta'$, and $\eta''$ are independent and
uniformly distributed on [0,1); for $|j-m|>1$, $x_j(1) = x_j(0)$.) 
At step 2 this process is repeated, with $m$ representing the site with the
global minimum of the variables $\{x_i(1)\}$, and so on.
In the {\it random neighbor} version of the model, the process is
realized on a complete graph (all sites are considered neighbors);
two randomly selected sites are updated in addition to the minimum $m$. 

We now define three modified Bak-Sneppen models, that
differ from the original only in the way that the
barriers $x_i$ evolve. In one, the site $M$ bearing the {\sl maximum}
of the $\{x_i\}$ is replaced with a random number $\eta$
and the two nearest neighbors are replaced with their own {\it square
roots}:
$x_M(t+1) = \eta$ and
$x_{M \pm 1}(t+1) = \sqrt{x_{M \pm 1}(t)}$.
We shall refer to this as the `radical' variant.
In the second variant,
the site with the maximum value is replaced by its own value {\it
squared},
while its two nearest neighbors receive random numbers $\eta$ and
$\eta'$:
$x_M(t+1) = [x_M(t)]^2$, $x_{M + 1}(t+1) = \eta$, and
$x_{M - 1}(t+1) = \eta'$.
This will be called the `centered square' version.  Finally, we
define a `peripheral square' variant, 
in which one of the nearest neighbors of $M$ is squared, while
$M$ and its other neighbor are replaced with random numbers:
$x_{M'}(t+1) = [x_{M'}(t)]^2$, 
$x_{M}(t+1) = \eta$ and $x_{M''}(t+1) =\eta '$.
(Here $M'=M+\sigma$ and $M''=M-\sigma$, where $\sigma$ is a random
variable that assumes values of +1 and -1 with equal probability.)

The motivation for studying these variants is twofold.  First, the 
Bak-Sneppen
model is notable for exhibiting a singular stationary probability density,
and it is of interest to examine the effect of changes in the dynamical rule
on this density and on the critical behavior.  If we introduce a
deterministic function $f(x)$ as part of the updating rule, it is desirable
that $f$ map the interval [0,1] to itself, making functions
of the form $f(x) = x^\alpha$ a natural choice.
In this context we note that the variants feature what may be
called `migration', that is, the systematic movement of certain variables
$x_i$ within the interval $[0,1)$.  In the radical variant the migration
is from the populated region (smaller $x$) to the `excluded' region
(larger $x$), whereas in the square versions migration occurs in the
opposite direction.

Secondly, the variants admit
interpretation as evolutionary processes.  (In the modified models,
we have for convenience defined the site with the maximum variable 
as the most vulnerable, so that small $x_i$ now corresponds to high 
fitness.) In the radical variant, replacement of the least-fit 
species provokes a reduction in the fitness of its neighbors, 
without leading to their immediate extinction.
Thus, some memory of the fitness of the neighboring species is retained.
The radical variant therefore seems a plausible modification
of the original model, in the biological context.  
In the peripheral square variant the extinction of the least-fit species
provokes extinction of one neighbor, and increased fitness of the other.
Finally, in the centered-square variant, 
$x_M \to x_M^2$ represents an increase in the
fitness of the least viable element of the system, 
while its neighbors go extinct.

\section{Mean-field theory}

\subsection{Original Model}

We develop mean-field approximations for the original and modified models,
along the lines of Refs. \cite{Dickman et al:2000} and \cite{head}.
To begin, we relax the extremal condition introducing a flipping rate of
$\Gamma e^{-\beta x_i}$ at site $i$, where $\Gamma ^{-1}$ is
a characteristic time, irrelevant to stationary properties, and
which we set equal to one ($\Gamma=1$).  Call this 
regularized system the `finite-temperature'
model. The extremal dynamics of
the original model is recovered in the zero-temperature limit, 
$\beta \rightarrow \infty$.

Consider the probability density $p(x)$.   
In the finite-temperature version of the 
original model, $p(x)$ 
satisfies

\begin{equation}
{dp(x,t)\over{dt}} = -e^{-\beta x} p(x,t) -2\int_0^1 e^{-\beta y} p(x,y,t)dy
+3\int_0^1 e^{-\beta y} p(y,t)dy ~~,
\label{eq:1}
\end{equation}

\noindent where $p(x,y,t)$ is the joint density for a pair of
nearest-neighbor sites and $p(y,t)$ is the one-site marginal density.
Invoking the mean-field factorization $p(x,y)=p(x)p(y)$ 
(we suppress the time argument from here on), we find:

\begin{equation}
{dp(x)\over{dt}} = -[e^{-\beta x} + 2I(\beta)]p(x) +3I(\beta) ~~,
\label{eq:2}
\end{equation}

\noindent where

\begin{equation}
I(\beta) \equiv \int_0^1 e^{-\beta y} p(y)dy ~~ ,
\label{eq:3}
\end{equation}
represents the overall flipping rate.  Eq. (\ref{eq:2}) is a nonlinear
equation for $p$, in which the density at $x$ is coupled to $p$ at
all other arguments via $I(\beta)$.  

In the stationary state we have
\begin{equation}
p(x) = \frac{3I}{2I+ e^{-\beta x}} ~~ .
\label{pss}
\end{equation}
Multiplying by $e^{-\beta x}$ and integrating over the 
range of $x$, we find 
\begin{equation}
I(\beta) = (e^{2\beta/3}-1)/[2e^\beta(1-e^{-\beta/3})], 
\end{equation}
and thus
\begin{equation}
p_{st}(x) = {3\over{2}}
{{1-e^{-2\beta/3}}\over{1-e^{-2\beta/3}+e^{-\beta x}(e^{\beta/3}-1)}}
~~.
\label{eq:4}
\end{equation}
This solution is plotted for various $\beta$ values in Fig. 1.

In the limit $\beta \rightarrow \infty$ the solution becomes a 
step function:

\begin{equation}
p_{st}(x) = {3\over{2}} \Theta(x-1/3)\Theta(1-x) ~~ .
\label{eq:5}
\end{equation}

\noindent Thus the mean-field approach correctly predicts a step-function
singularity for the probability density, although it places
the critical barrier at $x^* = 1/3$, 
whereas it actually falls at 0.66702(8) \cite{Grassberger:1995}.
On the other hand, the slightest relaxation of the extremal
condition destroys the singularity \cite{head}, since $p(x)$ is a smooth
function for $\beta < \infty$.  The rate of convergence to the step-function
is generally exponential with $\beta$, away from the discontinuity. 
The curves for various $\beta$ values exhibit an approximate
crossing near $x=1/3$. 
The derivative at this point however diverges only linearly with $\beta$:
$(dp_{st}/dx)_{x=1/3} \simeq 3\beta/8$ for large $\beta$.

Using Eq. (\ref{eq:2}), we find, in the limit of large $\beta$, 
that the relaxation time for a small disturbance
from the stationary solution grows $\sim  e^{\beta /3}$.
(By `small' we mean $I(\beta) \simeq e^{-\beta/3}/2$.)

The following observation will prove useful in the discussion
of the modified models.
If we assume that, in the limit $\beta \to \infty$, $p_{st}(x)$
is identically zero for $x < x^*$, and that $p_{st} \geq C > 0$
on some interval $[x^*, a]$ (in other words, the density
suffers a jump discontinuity at $x^*$), then
$I=\int_0^1 e^{-\beta y} p(y)dy \sim e^{-\beta x^*}$ and
so $e^{-\beta x}/I \sim e^{-\beta (x-x^*)}\rightarrow 0$ for
$x >x^*$. Then Eq. (\ref{pss}) reduces to the step-function expression,
Eq. (\ref{eq:5}).  Note however that 
$\lim_{\beta \to \infty} e^{\beta x^*} I(\beta) = 1/2$ not $3/2\beta$, 
as would be found by naively inserting the limiting density, Eq.(\ref{eq:5}),
in Eq.(\ref{eq:3}).  This means that the dominant contribution to
$I$ is due to the interval $[0,x^*]$, even in the limit $\beta \to \infty$,
which is readily seen if we write Eq.(\ref{eq:4}), for large $\beta$, as
\begin{equation}
p_{st}(x) \simeq \frac{3}{2}[ \Theta(x-x^*) + 
\Theta(x^*-x)e^{\beta (x-x^*)}]  \;.
\label{decomp}
\end{equation}
In the limit $\beta \to
\infty$, sites with $x < x^*$ constitute a set of probability zero,
but the site $m$ selected for extinction belongs to this set
{\it with probability one}.
This is a singular property of the Bak-Sneppen
model in the infinite-size limit, as discussed in the next subsection.  
(The infinite-size limit is implicit in mean-field theory.)

Although in the modified models we are unable
to find an analytical solution for finite $\beta$, it is possible to
integrate the mean-field equation numerically. 
Due to the factor $e^{\beta x}$, for large $\beta$, 
a very small time step would be needed to avoid instability
in the usual integration methods (e.g., Euler or Runge-Kutta).
We circumvent this difficulty using a partial integration
method \cite{dickman02}. To apply this method to the MF equation for
the original model, we write Eq. (\ref{eq:2}) 
in the form
\begin{equation}
{dp(x,t)\over{dt}} = -f(t)p(x,t)+g(t) ~~,
\label{eq:7b}
\end{equation}

\noindent where $f(t)=e^{-\beta x} + 2I(t)$ and $g(t)=3I(t)$. 
The formal solution is

\begin{equation}
p(x,t) =\exp[-\int_0^t du f(u)] \left\{ p(x,0) + \int_0^t dt' 
\exp \left[\int_0^{t'}
dt'' f(t'')\right] g(t')\right\} ~~.
\label{eq:7c}
\end{equation}

\noindent For a small time interval $\Delta t$, we find

\begin{eqnarray}
\nonumber
p(x,\Delta t) &\simeq& e^{-f(0)\Delta t} 
\left\{ p(x,0) + g(0) \int_0^{\Delta t} dt'
e^{f(0)t'}\right\} \\
&=&
e^{-f(0)\Delta t}p(x,0) + \frac{g(0)}{f(0)} (1-e^{-f(0)\Delta t}) ~~.
\label{eq:7d}
\end{eqnarray}

\noindent This relation can be iterated to find the
evolution of $p(x,t)$
from a given initial distribution, which converges to the stationary
density.

\subsection{BS model on a finite complete graph} 
\label{sec3B}

Mean-field theory is exact for the ``random neighbor" model, which may
also be thought of as the BS model on a complete graph, i.e., one
in which all pairs of sites are neighbors.  (When $x_m$ is updated,
two of these neighbors are chosen at random for updating as well.)
In this subsection we analyze the BS model with extremal dynamics
on a complete graph of $N$ sites.

Since sites are assigned independent
random numbers, the $x_i$ are independent, identically distributed
random variables drawn from the density $p(x,t)$.
Define the {\it distribution function} 
$P(x,t) = \int_0^x p(y,t) dy$.  The probability that the next site
to be updated, $x_m$, lies between zero and $x$ is:
\begin{equation}
\mbox{Prob}[x_m \leq x] = 1 - \left[ 1 - P(x) \right]^N  \;,
\label{prmleqx}
\end{equation}
i.e., one less the probability that the minimum is larger than $x$.
The probability that a randomly chosen neighbor has $x_i \leq x$
is simply $P(x)$, and the probability that one of the updated
sites receives a number $\leq x$ is $x$.  At each step, therefore,
the expected change in the number of sites with $x_i \leq x$ is
$3x - 2P(x) - \mbox{Prob}[x_m \leq x]$, which implies
\begin{equation}
\frac{dP(x,t)}{dt} = -\left\{1 - \left[ 1 - P(x) \right]^N\right\}
 -2P(x) + 3x        \;.
\label{eqmoP}
\end{equation}
(Here we have taken the time unit to represent $N$ updates.)

Letting $Q \equiv 1-P$, we have in the stationary state,
\begin{equation}
Q^N + 2Q - 3(1-x) = 0  .
\label{eqQ}
\end{equation}
(Note that $Q(0) = 1$, $Q(1) = 0$, and $dQ/dx \leq 0$.)
Numerical solution (Fig. 2) shows that for large $N$, $P_N(x)$ approaches
a singular function that is zero for $x < 1/3$, 
while for $x>1/3$, $P(x) = 3(x-1/3)/2$.  It is straightforward to
show that for fixed $x$ and $N \to \infty$,
\begin{equation}
P(x) \simeq \left\{
\begin{array}{l}
1 - (1 - 3x)^{1/N} \;,\;\;\;\; x < 1/3 \\
\; \\
\frac{3x-1}{2}  + \frac{1}{2} [\frac{3}{2}(1-x)]^N, \;\;\;\; x > 1/3 .
\end{array} 
\right.
\label{asymP}
\end{equation}
For $x = 1/3$ we have $\ln P \simeq -\ln N + \ln \ln (N/2)$ 
(plus terms of lower order in $N$) as $N \to \infty$.  Of interest is
the exponential convergence of $P$ to its limiting form
for $x > 1/3$, compared with algebraic convergence
for $x < 1/3$.  Note also that $\mbox{Prob}[x_m \leq 1/3]
\simeq 1 - 2/N$, so that the minimum indeed belongs
to the excluded region with probability one,
when $N \to \infty$.

\subsection{Pair approximation}

The analysis of the finite-temperature model
is readily extended to the pair level,
in which one studies the evolution of the two-site 
joint probability density $p(x,y)$.  Our starting point is
the following exact relation, obtained using
the same reasoning that led to Eq. (\ref{eq:1}):

\begin{eqnarray}
{dp(x,y)\over{dt}} &=& -\left(e^{-\beta x} + e^{-\beta y} \right) p(x,y) 
-\int_0^1 e^{-\beta u} \left[p(x,y,u) + p(u,x,y) \right] du
\nonumber
\\
&+&\int_0^1 \int_0^1 \left(e^{-\beta u} + e^{-\beta v} \right) p(u,v) dudv 
\nonumber
\\
&+& \int_0^1 \int_0^1 \left[p(x,u,v) + p(v,u,y) \right]
e^{-\beta v} dudv  \;.
\label{pair1}
\end{eqnarray}

\noindent Now invoking the pair factorization,
\begin{equation}
p(x,y,u) \simeq \frac{p(x,y)p(y,u)}{p(y)} ,
\label{pairfac}
\end{equation}
Eq. (\ref{pair1}) becomes
\begin{eqnarray}
{dp(x,y)\over{dt}} &=& -\left[e^{-\beta x} + e^{-\beta y} 
+ K(x) + K(y) \right] p(x,y)  +2I
\nonumber
\\
&+& \int_0^1 \left[p(x,u) + p(y,u) \right] K(u) du  \;,
\label{pair2}
\end{eqnarray}
with 
\begin{equation}
K(x) = J(x)/p(x),
\end{equation}
\begin{equation}
J(x) = \int_0^1 p(x,u) e^{-\beta u} du
\end{equation}
and 
\begin{equation}
I = \int_0^1 p(x) e^{-\beta x} dx = \int_0^1 J(x) dx \;.
\end{equation}

To find the stationary solution numerically, we note that
\begin{equation}
p(x,y) = \frac{2I + 
\int_0^1 \left[p(x,u) + p(y,u) \right] K(u) du}
{e^{-\beta x} + e^{-\beta y}  +K(x) +K(y)}.
\label{pair3}
\end{equation}
Starting from an arbitrary normalized density $p_0$ (for example,
uniform on $[0,1] \times [0,1]$), we generate $p_1$ by evaluating
the r.h.s. of Eq. (\ref{pair3}) using $p_0$ and normalizing the
resulting expression.  This procedure is then iterated until it
converges to the stationary density.  
We find that for large $\beta$, the stationary marginal density
approaches the step-function solution
\begin{equation}
p(x) = \frac{1}{1-x^*} \Theta(x - x^*) \Theta (1-x)
\label{pair4}
\end{equation}
with $x^* \simeq 0.47186$, a considerable improvement over the
site approximation.  In this limit, the joint distribution is 
the product of two identical one-site distributions.  Once again,
the portion of the unit square that has probability zero is in fact
responsable for all transitions.  In the region 
$D \equiv \{(x,y)|0 < x,y < x^*\}$, the
two variables are correlated, as shown by the nonzero correlation 
coefficient
$\rho \equiv \mbox{cov}(x,y)/\sqrt{\mbox{var}(x) \mbox{var}(y)}$.  
In the pair approximation,
$\rho \simeq 0.327$ for $(x,y) \in D$, as $\beta \to \infty$.

\subsection{Radical Variant}

In the radical model, the 
probability density $p(x)=p_X(x)$ satisfies:

\begin{eqnarray}
\frac{dp_X(x)}{dt} &=& -e^{\beta x} p_X(x) -2\int_0^1 e^{\beta y}
p_X(x,y)dy +\int_0^1 e^{\beta y} p_X(y)dy 
\nonumber
\\
&+& 2p_{X^{1/2}}(x)\int_0^1
e^{\beta y} p_X(y)dy ~~,
\label{eq:8}
\end{eqnarray}

\noindent where 
$p_{X^{1/2}}(x)=2xp_X(x^2)$.
With the definition $I(\beta) \equiv \int_0^1 e^{\beta y} p(y)dy$ and
the mean-field assumption
$p(x,y)=p(x)p(y)$, Eq. (\ref{eq:8}) reduces to

\begin{equation}
{dp(x)\over{dt}} = -e^{\beta x} p(x) + I(\beta)[-2p(x)+1+4xp(x^2)] ~~.
\label{eq:9}
\end{equation}

Given the step-function form of the stationary density for the original model,
it is reasonable to expect that in this case as well, $p(x)$ will have a
jump discontinuity for $\beta \to \infty$, and be zero for $x > x^*$.  We can get some insight 
into the nature of the density as $\beta \to \infty$ by first observing that
if $p(x) \leq C < \infty$ as $x \to 1$, then $I \sim Ce^{\beta}/\beta $ for large
$\beta$, and therefore $e^{-\beta} I \to 0$ as $\beta \to \infty$.  This in fact
holds unless $p(x)$ has a $\delta$-like contribution at $x=1$, which is not
expected since it is precisely the largest values of $x$ that are removed
in the dynamics.
Write the stationary solution to Eq. (\ref{eq:9}) as
\begin{equation}
p(x) = \frac{\overline{I}[1+4xp(x^2)]}{1+2\overline{I}} \;,
\label{radst}
\end{equation}
with $\overline{I} = e^{-\beta x} I$.  This gives $p(1) = 0$ in the
limit $\beta \to \infty$.  Similarly, supposing that $x p(x^2) \to 0$
as $x \to 0$, we have $p(0) = 1/2$ in this limit.  A similar line of
reasoning can be used to show that $dp/dx|_{x=1} = 0$ in the $\beta \to \infty$
limit.

The preceding discussion suggests that in the limit $\beta \to \infty$,
$p_{st}(x)$ is {\it identically zero} over some finite range $[x^*,1]$.
Assuming this to be so, we have $I \sim e^{\beta x^*}$
and in the limit $\beta \to \infty$ the stationary solution is given by the 
{\it functional equation:}

\begin{equation}
2p(x)-4xp(x^2)=1 \;,
\label{eq:10}
\end{equation}
for $0 \leq x < x^*$.
Writing this as $p(x)=1/2+2xp(x^2)$, we can iterate 
to find $p(x^2)=1/2+2x^2p(x^4)$,
$p(x^4)=1/2+2x^4p(x^8)$,
$p(x^8)=1/2+2x^8p(x^{16})$, etc., which suggests the solution

\begin{equation}
p_1(x)={1 \over 2} + x + 2x^3 + 4x^7 + 8x^{15} + ... = \sum^\infty_{i=0}
2^{i-1} x^{2^i-1} ~~.
\label{eq:11}
\end{equation}

\noindent Substituting this `lacunary series' in Eq.(\ref{eq:10}), 
one readily verifies that $p_1(x)$ is a solution.
Similarly, rewriting Eq.(\ref{eq:10}) as $p(x^2)=-1/4x+p(x)/2x$,
one finds $p(x)=-1/4x^{1/2}+p(x^{1/2})/2x^{1/2}$,
$p(x^{1/2})=-1/4x^{1/4}+p(x^{1/4})/2x^{1/4}$, etc., leading to a second
solution:

\begin{equation}
p_2(x)= - {1 \over {4x^{1/2}}} - {1 \over {8x^{3/4}}} - {1 \over
{16x^{7/8}}} - {1 \over {32x^{15/16}}} - ... =
- \sum^\infty_{i=1} 2^{-(i+1)} x^{-(1-2^{-i})} ~~.
\label{eq:12}
\end{equation}

We now search a solution of the form $p(x)=Ap_1(x)+Bp_2(x)$;
substituting in Eq. (\ref{eq:10}), yields the condition
$A+B=1$. This linear combination must 
however be normalizable.  The relevant integrals are:

\begin{equation}
\int^{x^*}_0 p_1(x) dx = {1 \over 2} \sum^{\infty}_{n=0} {x^*}^{2^n}= {1
\over 2} (x^*+{x^*}^2+{x^*}^4+{x^*}^8+...) ~~,
\label{eq:13}
\end{equation}

\begin{equation}
\int^{x^*}_0 p_2(x) dx = - {1 \over 2} \sum^{\infty}_{n=1} {x^*}^{2^{-n}} 
= -{1 \over 2} (1+{x^*}^{1/2} +{x^*}^{1/4} +{x^*}^{1/8}+...) ~~.
\label{eq:14}
\end{equation}
Since $0<x^*<1$, the first integral converges while the second diverges, so that
$p_2(x)$ is not normalizable, implying $B=0$.  A normalized, positive
solution is
$p(x)=p_1(x)$ for $x < x^* =0.793189$
and $p(x) = 0$ for $x > x^*$.  ($x^*$ is determined by
normalization.)
This solution, for infinite $\beta$, is plotted in Fig. 3.
Numerical integration of Eq. (\ref{eq:9}) through the method outlined
in Sec. IIIA yields results consistent with this solution, 
as may again be seen in the figure. Finally, simulation of the 
random-neighbor version of the model yields a stationary 
distribution consistent with this expression (see Fig. 4).

We will now analyze the radical model on a finite complete graph
and show that its probability density (equation \ref{eq:11})
can be derived via a different path.
Define the distribution function $Q(x,t)=\int_{x}^{1} p(y,t) dy$.
By the same reasoning developed in section \ref{sec3B}, the probability
that $x_m$, the next site to be updated, lies between $x$ and 1 is
\begin{equation}
\mbox{Prob}[x_m \geq x] = 1 - \left[ 1 - Q(x) \right]^N  ~~.
\end{equation}
The probability that a randomly chosen neighbor has $x_i \geq x$ is 
$Q(x)$, while the probability that a neighbor receives a barrier
$\geq x$ is $Q(x^2)$. The probability that $x_m$ receives a new value
between $x$ and 1 is simply $1-x$. The expected change in the
number of sites with $x_i \geq x$ is 
$(1-x)+2Q(x^2)-2Q(x)-\{1 - \left[ 1 - Q(x) \right]^N\}$,
so that
\begin{equation}
\frac{dQ(x,t)}{dt} = (1-x)+2Q(x^2)-2Q(x) -\left\{1 - \left[ 1 - Q(x) \right]^N\right\} ~~.
\label{eq:radicalcg}
\end{equation}

In the sationary state, the definition $P \equiv 1-Q$ leads to 
\begin{equation}
x+2P(x^2)-2P(x)-P(x)^N = 0 ~~.
\end{equation}
(Note that $P(0)=0$ and $P(1)=1$). Since $0\leq P(x) < 1$ for $x<x^*$,
$\lim_{N\to \infty} P(x)^N = 0$ and the probability density in an infinite
system obeys the functional equation $x+2P(x^2)-2P(x)=0$. Using the 
iterative
approach, we find two solutions, one divergent (which is rejected), 
and the finite solution
\begin{equation}
P(x)={1\over 2}\sum_{n=0}^{\infty} x^{2^n} ~~ ,
\end{equation}
\noindent 
i.e., the integral of $p_1(x)$, Eq. (\ref{eq:11}).
Normalization then demands that $P = 1$ for
$x\geq x^*$.

Numerical solution of equation (\ref{eq:radicalcg}) (see Fig. 5) shows that,
for large $N$, $P_N(x)$ approaches the singular function described above. 
For fixed $x> {x^{*}}^{1/2}$, and $N\to \infty$,
we find $P(x) \simeq 1 - x^{1/N} $.

\subsection{Centered Square Variant}

The probability density in the finite-temperature version of the
centered square variant obeys:
\begin{equation}
{dp_X(x)\over{dt}} = -e^{\beta x} p_X(x) -2\int_0^1 e^{\beta y}
p_X(x,y)dy +2\int_0^1 e^{\beta y} p_X(y)dy + {e^{\beta 
\sqrt{x}}\over{2\sqrt{x}}} p_X(\sqrt{x}) ~~.
\end{equation}
Mean-field factorization leads to
\begin{equation}
{dp(x)\over{dt}} = -e^{\beta x} p(x) -2I(\beta)[p(x)-1] + {e^{\beta 
\sqrt{x}}\over{2\sqrt{x}}} p(\sqrt{x}) ~~,
\label{aqui1}
\end{equation}
with $I(\beta)$ as defined in Sec. IIID. In the stationary state, we 
find
\begin{equation}
p(x) = { { 2 + {1\over {2\sqrt{x}}} p(\sqrt{x}) e^{\beta \sqrt{x}}/I 
}\over{2+e^{\beta x}/I} } ~~ .
\label{pcsv}
\end{equation}
The hypothesis that in the limit $\beta \rightarrow \infty$ ,
the stationary density $p(x)=0$ for $x>x^{*}$ 
implies $I(\beta)\sim e^{\beta x^*}$. Therefore
\begin{equation}
\lim_{\beta \rightarrow \infty} {e^{\beta x}\over I} =
\left\{ 
\begin{array}{c}
0 \ \ \ if \ \ \ x<x^{*} \\ 
\infty \ \ \ if \ \ \ x>x^{*} 
\end{array}
\right.  
\end{equation}

\noindent and

\begin{equation}
\lim_{\beta \rightarrow \infty} {e^{\beta \sqrt{x}}\over I} =
\left\{ 
\begin{array}{c}
0 \ \ \ if \ \ \ x<{x^*}^2 \\ 
\infty \ \ \ if \ \ \ x>{x^*}^2 
\end{array}
\right. 
\end{equation}

This, combined with Eq. (\ref{pcsv}),  
implies that $p(x)=1$ for $x \in [0,{x^*}^2]$, and that 

\begin{equation}
p(x) = \lim_{\beta \rightarrow \infty} {\left[1+{1\over{4 \sqrt{x}}} 
{e^{\beta \sqrt{x}}\over I} p(\sqrt{x})\right]}  \ \ \ {\rm for} \ x \in 
[{x^*}^2,x^*]
\label{aqui2}
\end{equation}

\noindent The iterative process used in the
preceding section
is not useful here as it leads to pathological solutions, due to the 
divergent ratio ${e^{\beta \sqrt{x}}/ I}$ in this interval.  
Observe that if, for $x \in [x^*,x^{* 1/2}]$,
$p(x)= g(x^2){I / e^{\beta x}}$, as
$\beta \to \infty$, with $g(x^2)$ finite, then
$p(x)=1+g(x)/4\sqrt{x}$ for $x \in [{x^*}^2,x^*]$. (Note that 
this means that $p(x)\to 0$ as $\beta \to \infty$ for $x$
in the interval $[x^*,{x^*}^{1/2}]$,
consistent with the hypothesis that $p(x) \to 0$ for $x>x^*$.)
The function $g(x)$ is however yet to be determined. 
Attempting the simplest solution, $g(x)=$ constant, we find a surprisingly
reasonable result, as shown in Fig. 6. Next, allowing
$g(x)=ax+b$, with $a$ and $b$ constant, yields excellent
agreement with the numerical integration of Eq. (\ref{aqui1})
as also shown in Fig. 6.  We do not have
an argument why $g(x)$ should take this form. 

The threshold $x^*$ can be determined in a simple way. 
Since the fraction of barriers in the interval $[x^*,1]$
is constant in the stationary state, the mean number of barriers 
removed from this interval at each time step
must equal the mean number inserted.
The probability that the maximum $x_M$
lies in $[x^*,1]$ is 1, while its random neighbors are
certainly below $x^*$. Each updated neighbor
has a probability
$(1-x^*)$ of receiving a barrier in the interval $[x^*,1]$,
while the maximum remains in $[x^*,1]$ with 
probability $p_1$, the probability that 
$x_M \in [{x^*}^{1/2},1]$. Thus, we have

\begin{equation}
1=2(1-x^*)+p_1 ~~.
\label{aqui3}
\end{equation}

This reasoning can be repeated for the intervals $[{x^*}^{1/2},1]$,
$[{x^*}^{1/4},1]$, ..., leading to

\begin{eqnarray}
p_1 = 2(1-{x^*}^{1/2}) + p_2 ~~ , \\
p_2 = 2(1-{x^*}^{1/4}) + p_3 ~~~~ ...
\end{eqnarray}

\noindent where $p_n$ is the probability
that $x_M \in [{x^*}^{1/2^n},1]$.
Substituting this result in equation (\ref{aqui3}), 
we find 

\begin{equation}
1=2(1-x^*)+ 2(1-{x^*}^{1/2})+2(1-{x^*}^{1/4})+2(1-{x^*}^{1/8})+...~~,
\end{equation}

\noindent which provides $x^*=0.761072$. Finally, we note that
normalization implies 
$\int_{{x^*}^2}^{x^*} g(x)/4\sqrt{x} ~ dx = 1-x^* = 0.238928$, 
providing a constraint on the function $g(x)$.

\subsection{Peripheral Square Variant}

We now apply the mean-field analysis to the peripheral square variant.
In this case,
the probability density satisfies:

\begin{eqnarray}
{dp_X(x)\over{dt}} &=& -e^{\beta x} p_X(x) -2\int_0^1 e^{\beta y}
p_X(x,y)dy +2\int_0^1 e^{\beta y} p_X(y)dy 
\nonumber
\\
&+& p_{X^2}(x)\int_0^1 e^{\beta
y} p_X(y)dy ~~,
\label{eq:15}
\end{eqnarray}

\noindent where 
$p_{X^2}(x)=p_X(x^{1/2})/2x^{1/2}$.
Under the mean-field factorization 
this reduces to

\begin{equation}
{dp(x)\over{dt}} = -e^{\beta x} p(x) + I(\beta)[-2p(x)+2+{1 \over
{2x^{1/2}}}p(x^{1/2})] ~~,
\label{eq:16}
\end{equation}
with $I(\beta)$ as given in Sec. IIIC.
The hypothesis that, for $\beta
\rightarrow \infty$, the stationary density
$p(x) = 0$ for $x > x^*$ implies $I(\beta)\sim e^{\beta x^*}$,
 leading to the functional equation:

\begin{equation}
p(x)-{1 \over {4x^{1/2}}}p(x^{1/2})=1 ~~~~~(\beta \rightarrow \infty)~~,
\label{eq:17}
\end{equation}
for $x < x^*$.
 
As in the centered-square variant, the iterative method does not
yield a useful solution, and we pursue a different approach.
Let $p(x)=f(x)$ on the interval ${x^*}^2 \leq x < x^*$, where the function $f(x)$
and the constant $x^*$ are yet to be determined. Using Eq. (\ref{eq:17}), we find that

\begin{eqnarray}
p(x)=1+{f(x^{1/2})\over
{4x^{1/2}}}                                              ~~ & {\rm for}
~~ {x^*}^4 \leq x < {x^*}^2 ~~,& \label{eq22}\\
p(x)=1+{1\over {4x^{1/2}}}+{f(x^{1/4})\over
{4^2x^{3/4}}}                        ~~ & {\rm for} ~~ {x^*}^8 \leq x < {x^*}^4
~~,& \label{eq23}\\
p(x)=1+{1\over {4x^{1/2}}}+{1\over {4^2x^{3/4}}}+{f(x^{1/8})\over
{4^3x^{7/8}}}  ~~ & {\rm for} ~~ {x^*}^{16} \leq x < {x^*}^8 ~~,& ~{\rm
etc.}\label{eq24}
\end{eqnarray}
Thus we have found a family of solutions $p(x)$ to Eq.(\ref{eq:17}),
which is quite general since $f(x)$ is still undetermined.

We now show that $f(x)=1$ on the interval $[x^{*2}, x^*)$.
To begin, we note that in the stationary state, Eq. (\ref{eq:16}) implies
\begin{equation}
\left(\frac{e^{\beta x}}{I} + 2 \right) p(x) = 
2 + \frac{p(\sqrt{x})}{2\sqrt{x}}  \;.
\label{ssrel}
\end{equation}
In particular, for $x=1$ we have
\begin{equation}
\left(\frac{e^{\beta}}{I} + \frac{3}{2} \right) p(1) = 
2  \;,
\label{ssrel1}
\end{equation}
so that if $I \simeq A e^{\beta x^*}$, then
$p(1) \simeq 2A e^{-\beta(1-x^*)}$.
It is readily seen that $p(x) \simeq 2A e^{-\beta(x-x^*)}$
satisfies Eq. (\ref{ssrel}) for $x^* < x \leq 1$.  The same
equation then leads to $p(x) = 1$ as $\beta \to \infty$,
for $x^{*2} < x < x^*$.
We may then develop the full solution using Eqs. (\ref{eq22}) to (\ref{eq24});
normalization implies $x^*=1/2$.
The result is the function $p(x)$ plotted in Fig. 7, which is in good
agreement with the density found via simulation of the random-neighbor model.
This solution is discontinous at $x^*$, $x^{*2}$, $x^{*4}$,..., 
and exhibits an integrable divergence at $x = 0$.  

The value of $x^*$ can be confirmed through the reasoning developed
in previous subsection for the peripheral square model, which in this case
implies $1 = 2(1-x^*)$ so that $x^*=1/2$.

\section{Numerical simulation and Critical exponents}

We now compare the mean-field theory predictions with
simulation results.  We estimate the probability density $p(x)$ on the
basis of a histogram of barrier frequencies, dividing 
[0,1] into 100 subintervals.
Histograms are accumulated after $N_{st}$ time steps, as
required for the system 
(a ring of $N$ sites) to relax to the stationary state. 
In order to improve statistics, we average over $N_r$ realizations.

The simulation results for the original BS model are shown in Fig. 8.
We observe qualitative agreement between nearest-neighbor (NN)
and random-neighbor (RN) versions.  
(The simulation parameters are: 
$N=1000$, $N_{st}=10^6$, $N_r=10^3$ (NN);
$N=2000$, $N_{st}=10^5$, $N_r=10^3$ (RN).)
Here and
in all other cases, the simulation result for the random neighbor
model appears to converge (as expected) to the mean-field
prediction.  In light of the discussion in Sec. IIIB, it is
reasonable to regard the rounding of the step function 
as a finite-size effect.

Fig. 4 presents similar results for the radical variant.
We notice that the NN and RN
versions exhibit qualitatively similar probability
densities, differing mainly in the value of
the threshold $x^*$.  (In this case we use
$N_{st}=10^6$ and $N_r=10^3$, with system sizes $N=1000$ (NN)
and 2000 (RN).)

Simulation results for the centered and peripheral square versions are shown
in Fig. 9 and 10, resp.. For the centered square,
the NN and RN,
probability densities are quite similar, differing mainly in the value of the
threshold and in the inclination of the central portion (Fig. 9).
(The simulation parameters are
$N=2000$, $N_{st}=10^7$, $N_r=500$ (NN),
$N=10000$, $N_{st}=10^6$, $N_r=10^3$ (RN).)
On the other hand, for the peripheral square version, the
nearest-neighbor 
and random-neighbor 
densities are somewhat different, since the latter exhibits
various steps, while only one such step is evident in the former 
(see Fig. 10).  Further study is needed to determine 
whether this is represents 
a qualitative difference between the two formulations, or is instead 
due to finite size and/or finite numerical resolution.  
(Note that for the square models we use larger lattices,
which were necessary to observe clear singularities.
The simulation parameters are:
$N=1000$, $N_{st}=10^6$, $N_r=10^3$ (NN),
$N=50000$, $N_{st}=10^7$, $N_r=20$ (RN).)

Several quantities are known to display power-law behaviour in the
Bak-Sneppen model \cite{Bak and Sneppen:1993,Grassberger:1995,Head:1998}.
In particular, we studied the distribution $P_J(r)$ that
sucessive updated sites are separated by a distance $r$.
In the original model, $P_J(r) \sim r^{-\pi}$, 
with $\pi=3.23(2)$ \cite{Head:1998}.  
(Figures in parentheses denote uncertainties.)
We performed simulations of the modified models using
$10^9$ time steps on lattices of 2000 or more sites, yielding (see Fig. 11),
$\pi=3.22(2)$ for the original BS model; 
$\pi=3.27(2)$ for the radical variant;
$\pi=3.25(2)$ and $3.24(2)$ 
for the centered and peripheral square variants,
respectively.
Thus we find strong evidence that all of the variants introduced here
belong to the same universality class as the original model.

\section{Conclusions}

In order to understand the implications of extremal dynamics, 
we propose several modified Bak-Sneppen
models. Although different updating rules lead to completely
different probability densities, they are always singular at one or
more points. The step-like singularities appear in the limiting densities,
either as $\beta \to \infty$ in the finite temperature model
(on an infinite lattice), or as the system size $N \to \infty$,
under extremal dynamics.  Thus the double limit of infinite size
and zero temperature is required for the BS model or it variants to
generate a singular probability density.

A remarkable feature common to the original model and all the variants
considered, is that in the infinite-size limit, and under extremal
dynamics, a certain interval $D \subset [0,1]$ has probability
zero, and yet contains the extremal site with probability one.
If we regard the density of active sites, $\mbox{Prob}[x \in D]$, as
the order parameter $\rho$, then the BS model and its variants are
seen to realize the `SOC limit' \cite{Dickman et al:2000,head}
$\rho \to 0^+$.  Correlations between site variables 
$x_i, x_j, ..., x_n$ are zero unless one or more of these values
falls in $D$.

Our conclusions regarding the form of the stationary densities are based 
on mean-field analyses that are exact for the random-neighbor versions,
as is verified numerically. We also present a pair approximation for the
original model. An important point is that mean-field theory captures the
form of the probability density correctly, as shown via
simulations of the modified models on a one-dimensional lattice.  
The latter suggest that the critical exponents are independent of dynamics.

The Bak-Sneppen model appears to be a
prototype for a large universality class, since countless variants,
beyond those presented here, are possible.  We expect any
dynamics that respects the symmetry of the original model (that is,
spatial isotropy), and that does not introduce new conserved 
quantities, to have the same exponents as the original model.
This is interesting, since the same critical behavior may subsist,
as it were, upon stationary distributions of very different forms.
Aside from its intrinsic interest, the question of 
universality is important for applications, since
the precise form of the dynamics in a specific setting
(e.g., evolution) is generally unknown, and probably 
quite different from that of the original model.  Power laws and a 
singular stationary density only appear in the extremal dynamics limit,
which may be difficult to realize in spatially extended natural systems.
\vspace{1em}

\bigskip
\noindent {\bf ACKNOWLEDGMENTS}
\smallskip

\noindent We thank CNPq and CAPES, Brazil,
for finacial suport.

\newpage

\begin{figure}
\epsfysize=6cm
\epsfxsize=6cm
\centerline{
\epsfbox{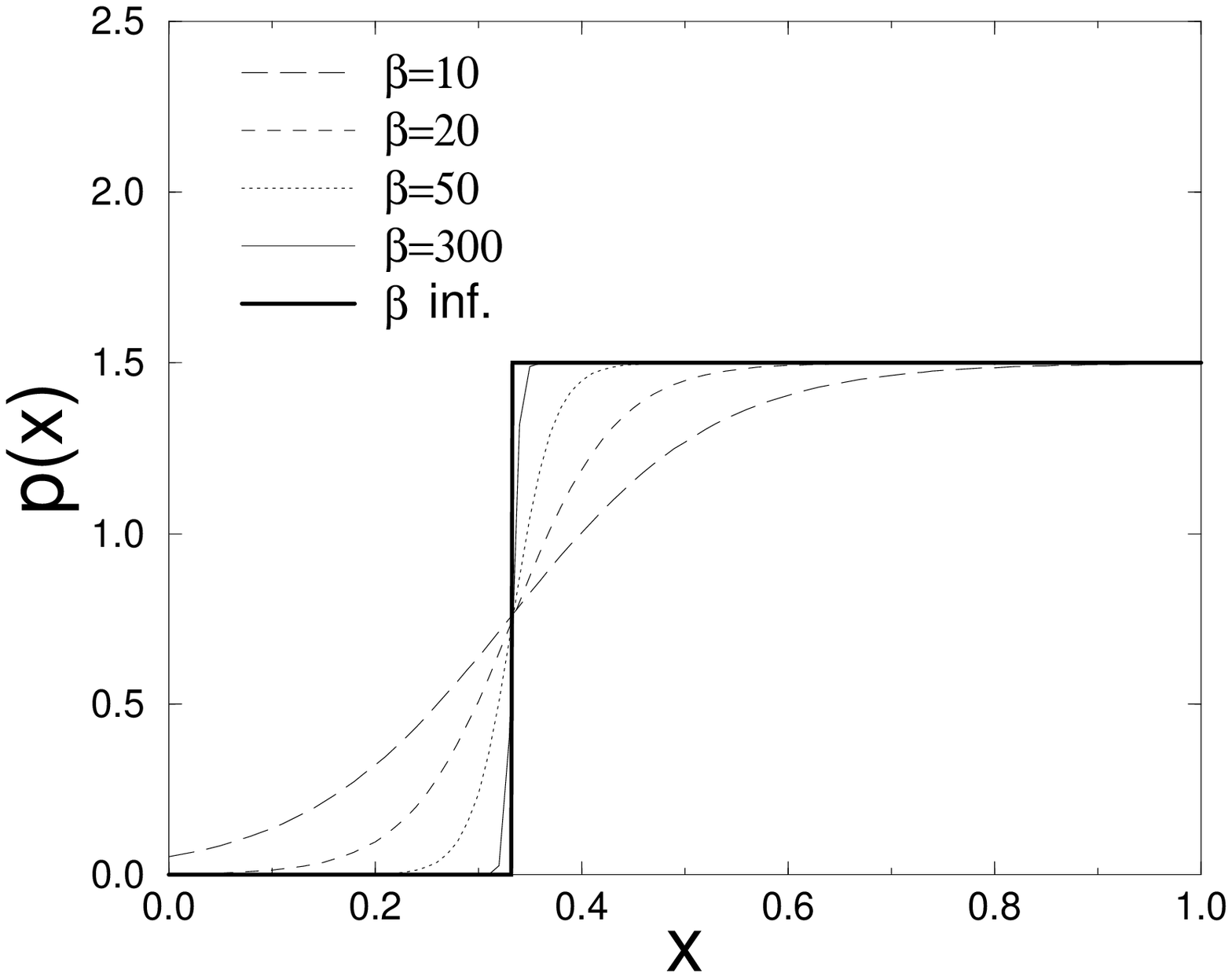}}
\caption{Original model: finite-temperature mean-field theory Eq. (6) for 
$\beta$ values as indicated ($\beta$ inf. stands for $\beta \to \infty$).}
\label{fig1}
\end{figure}

\begin{figure}
\epsfysize=6cm
\epsfxsize=6cm
\centerline{
\epsfbox{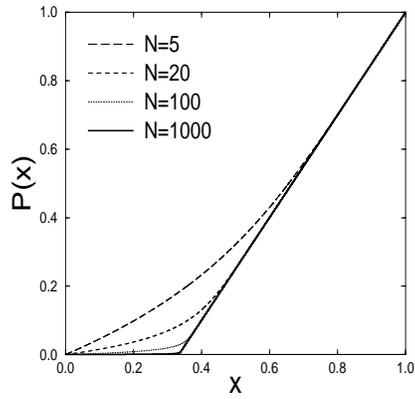}}
\caption{Original model on complete graph (extremal dynamics) for
system sizes as indicated.}
\label{fig0}
\end{figure}

\begin{figure}
\epsfysize=6cm
\epsfxsize=6cm
\vspace{-1cm}
\centerline{
\epsfbox{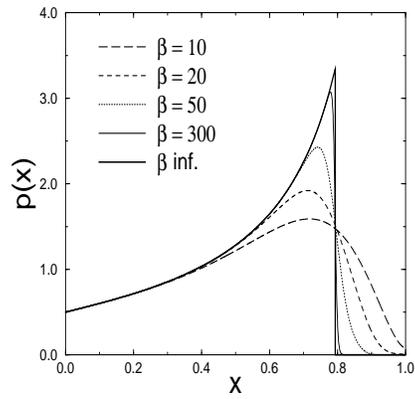}}
\caption{Radical variant: finite-temperature mean-field theory for
$\beta$ values as indicated.}
\label{fig3}
\end{figure}

\begin{figure}
\epsfysize=6cm
\epsfxsize=6cm
\centerline{
\epsfbox{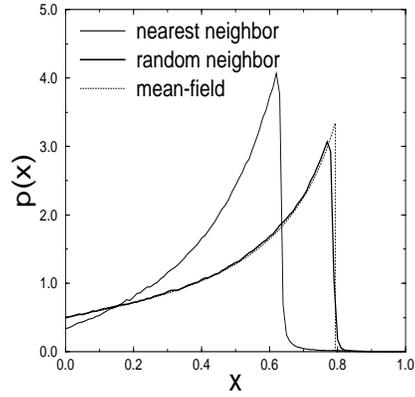}}
\caption{Radical variant: simulation results for nearest neighbor
 and random neighbor versions
compared with mean-field prediction.}
\label{fig4}
\end{figure}

\begin{figure}
\epsfysize=6cm
\epsfxsize=6cm
\centerline{
\epsfbox{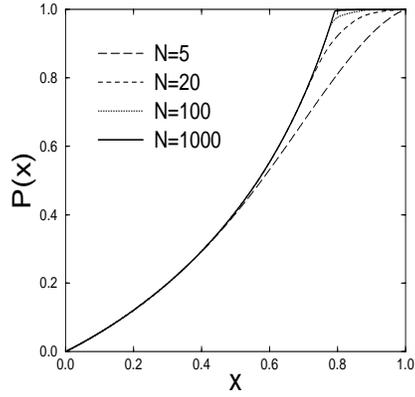}}
\caption{Radical variant on complete graph for system sizes as
indicated.}
\label{fig4b}
\end{figure}

\begin{figure}
\epsfysize=6cm
\epsfxsize=6cm
\centerline{
\epsfbox{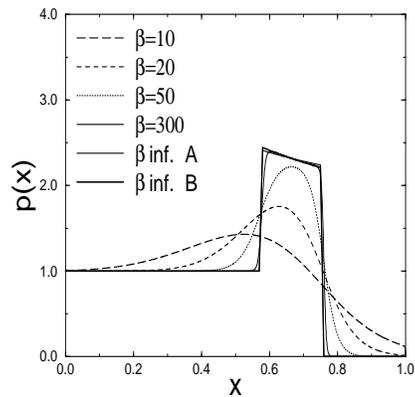}}
\caption{Centered square variant: finite-temperature mean-field theory for
$\beta$ values as indicated.  Approximation A refers to $g(x) =$ constant,
B for $g(x)=ax+b$, both at zero temperature, as explained in text.}
\label{fig5}
\end{figure}

\begin{figure}
\epsfysize=6cm
\epsfxsize=6cm
\centerline{
\epsfbox{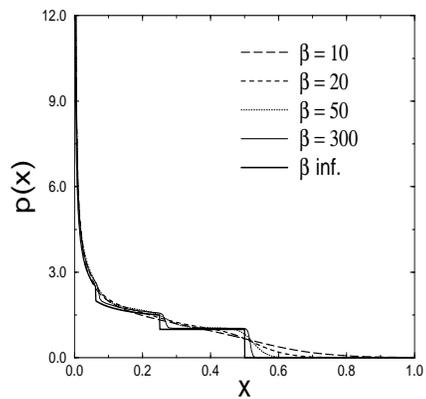}}
\caption{Peripheral square variant: finite-temperature mean-field theory for
$\beta$ values as indicated.}
\label{fig7}
\end{figure}

\begin{figure}
\epsfysize=6cm
\epsfxsize=6cm
\centerline{
\epsfbox{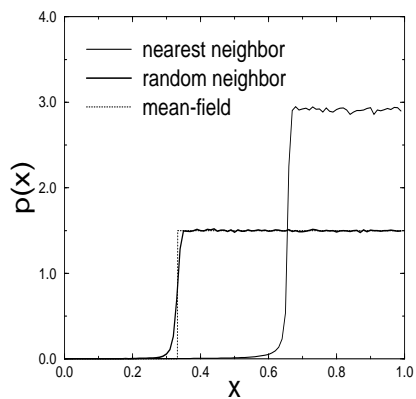}}
\caption{Original model: simulation results for NN and RN versions
compared with mean-field prediction.}
\label{fig2}
\end{figure}

\begin{figure}
\epsfysize=6cm
\epsfxsize=6cm
\centerline{
\epsfbox{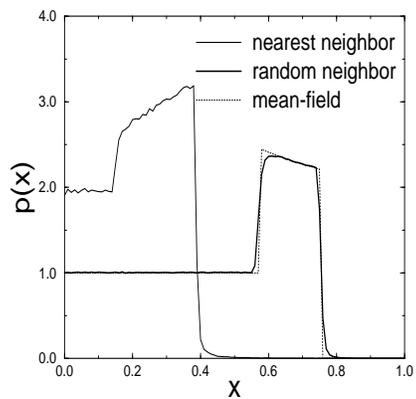}}
\caption{Centered square variant: simulation results for NN and RN versions
compared with mean-field prediction.}
\label{fig6}
\end{figure}

\begin{figure}
\epsfysize=6cm
\epsfxsize=6cm
\centerline{
\epsfbox{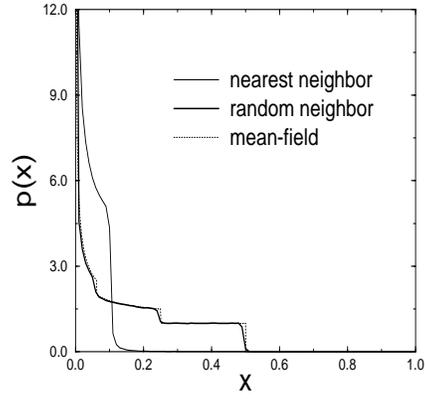}}
\caption{Peripheral square variant: simulation results for NN and RN versions
compared with mean-field prediction.}
\label{fig8}
\end{figure}

\begin{figure}
\epsfysize=6cm
\epsfxsize=6cm
\centerline{
\epsfbox{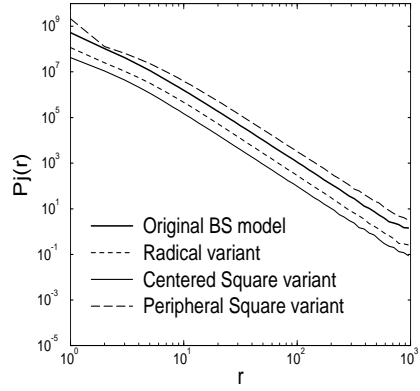}}
\caption{Distribution $P_J(r)$ of the distance $r$ separating sucessive updated sites.
The curves have been shifted vertically to facilitate comparison.}
\label{fig9}
\end{figure}

\end{document}